# Simulation of Patient Flow in Multiple Healthcare Units using Process and Data Mining Techniques for Model Identification

*Sergey V. Kovalchuk*[1], *Anastasia A. Funkner*[1], *Oleg G. Metsker*[1], *Aleksey N. Yakovlev*[1,2]

[1] ITMO University, Saint Petersburg, Russia

[2] Almazov National Medical Research Centre, Saint Petersburg, Russia

kovalchuk@mail.ifmo.ru, funkner.anastasia@gmail.com, olegmetsker@gmail.com, yakovlev_an@almazovcentre.ru

**Abstract.** ***Introduction:*** An approach to building a hybrid simulation of patient flow is introduced with a combination of data-driven methods for automation of model identification. The approach is described with a conceptual framework and basic methods for combination of different techniques. The implementation of the proposed approach for simulation of the acute coronary syndrome (ACS) was developed and used in an experimental study. ***Methods:*** Combination of data, text, and process mining techniques and machine learning approaches for analysis of electronic health records (EHRs) with discrete-event simulation (DES) and queueing theory for simulation of patient flow was proposed. The performed analysis of EHRs for ACS patients enable identification of several classes of clinical pathways (CPs) which were used to implement a more realistic simulation of the patient flow. The developed solution was implemented using Python libraries (SimPy, SciPy, and others). ***Results:*** The proposed approach enables more realistic and detailed simulation of the patient flow within a group of related departments. Experimental study shows that the improved simulation of patient length of stay for ACS patient flow obtained from EHRs in Almazov National Medical Research Centre in Saint Petersburg, Russia. ***Conclusion:*** The proposed approach, methods, and solutions provide a conceptual, methodological, and programming framework for implementation of simulation of complex and diverse scenarios within a flow of patients for different purposes: decision making, training, management optimization, and others.

## 1 Introduction

Detailed simulation of patients' care provided in healthcare units requires analysis of this process from multiple points of view. The process involves multiple departments, many types of activities, different actors with particular roles, expertise level, knowledge, etc. Moreover, multiple scopes and aspects of the healthcare process can be considered (department’s resources load, accounting in a hospital, patient’s personalized experience during the care process, and many others). One of the key reason of growing interest to this subject is the latest attention to patient-centered healthcare, including personalized medicine [1], value-based healthcare [2]. In contrast to 'traditional' well-grounded evidence-based medicine dealing with typical patients, these areas consider the diversity of patients with more attention taking care of personal combination of factors specific to the particular patient. This leads to certain interest to comprehensive simulation approaches [3,4], as well as automatic or semi-automatic ways to the analysis of medical records using machine learning [5], data mining [6], process mining [7] techniques, interaction with patients [8], and other approaches. Still, there is multitude of problems that reveal multiple sources of uncertainty in complex simulation tasks. The list of problems includes lack of consistency, completeness, and correctness of medical data to be analyzed [9], low coverage of rare cases with clinical pathways (CP) [10], weak formalization and high uncertainty in core medical knowledge [11]. To overcome these issues, we developed a hybrid approach introduced in the present work where a combination of techniques from data, process and text mining is proposed to support computer-aided simulation. The presented approach is aimed towards automation of comprehensive model and scenarios identification, classification of patient-centered CPs, and management of simulation applications.

The remaining of the paper is as follows. Section 2 presents a brief analysis of the area of modeling and simulation of patient flow in healthcare. Section 3 discusses the essential requirements identified for the developed approach. Section 4 introduces the basics of the conceptual framework developed to support simulation models building using electronic health records (EHR) and other information sources. Section 5 uncovers the details of the proposed approach showing the way of different methods are combined within a solution. Section 6 demonstrates an example of the proposed approach's application within a task of simulation of key departments involved in acute coronary syndrome (ACS) treatment procedures performed in Almazov National Medical Research Center (Almazov Centre)[1] (Saint Petersburg, Russia), one of the leading cardiological centers in Russia. Finally, the last section presents concluding remarks and future development of the proposed ideas.

## 2 Related works

### 2.1 Simulation

Simulation is a widely spread technique for investigation in medicine and healthcare (e.g., see a survey on multi-department simulation [3]). Applications and goals of simulation vary in different projects: analysis of selected department load [4,12], department management [13,14], policy making [15,16], building testbeds for testing of various solutions [17], etc. Most popular approaches are based on discrete-event simulation (DES) [12,15], system dynamics (SD) [8,18], agent-based modeling (ABM) [19], queueing theory [13,20]. Although there is a multitude of works in the area, there still are unresolved issues. For example, authors of work [3] mention that, while there are multiple works on simulation of selected departments (mainly, intensive care or emergency departments), works focused on simulation with coverage of multiple departments within complex scenarios is presented weakly. Another problem, attracting the attention of the researchers is a combination of various models within a single solution. E.g., authors of [4] try to combine ABM, SD, and DES approaches within a single multiscale solution.

[1] http://www.almazovcentre.ru/?lang=en

### 2.2 Clinical pathways

One of the important issue for simulation of clinical departments and patient flow through these departments is the identification of "trajectories" which patients follow during the treatment process. This idea is tightly connected to the concept of a clinical pathway (CP) which define typical care paths for certain group of patients. This concept is widely spread, but still, it has enormous variation in its definition mentioning different techniques, goals, and characteristics (see [21,22] demonstrating attempts to unify this concept). From the simulation point of view, CPs include important information on main steps and procedures applied during care process. Commonly, CPs are developed manually by a group of experts and physicians to define explicitly key elements of care process based on guidelines, best practices, and available experience. Still, having high variation an uncertainty in patient flow, automatic identification of CPs become a significant issue within a context of personalization in medicine and healthcare. Popular solutions for automatic identification of CPs are based on process mining (PM) [7] and data mining (DM) [6] techniques which gain popularity in medicine and healthcare applications. For example, authors of [23] use PM for identification of rare CPs as a more productive alternative to expert-based CP development. In [24] a PM approach to CPs analysis is aimed towards fraud detection in healthcare. Work [25] presents an extension of PM approaches with composite state machines to model complex processes. A significant impact in CPs identification and analysis may be achieved with the introduction of DM techniques. E.g., in [26] authors use DM to identify conditional branching in process models, [27] presents DM and graph analysis applied to the investigation of temporal patterns in CPs, [28] introduces an original approach based on Latent Dirichlet Allocation within a combination of PM and DM approaches. An important problem attracting the attention of the researchers is clustering and classification of patients for identification of different CPs in a selected group of patients [29,30].

### 2.3 Predictive modeling

Another important area of DM application is predictive modeling in medicine and healthcare [31]. The area includes a multitude of tasks being solved within various approaches. For instance, an important task for policy making, resource management and utilization, decision support, planning, and optimization (in terms of cost, effectiveness, etc.) in healthcare is prediction of length of stay (LoS) for various patients which can be solved with multiple techniques from DM area [32–38]. Another important task is the prediction of clinical outcomes. For example, Asadi et al. [39] used several machine learning techniques to predict potential outcomes (good and poor) of patients with acute ischemic stroke. Oermann et al. [40] produce a similar study for patients with stereotactic radiosurgery for cerebral arteriovenous malformations. Copper et al. tried 11 statistical and machine learning techniques to predict dire outcomes for patients with community-acquired pneumonia [41]. Extending this task multiple researchers tries to predict complications, comorbidities, severity of a disease. E.g. DM is used to predict short-term complications of type 2 diabetes mellitus [42], long-term complications after coronary surgery [43]. SVM, linear regression, decision tree, and forest are the most common methods for predicting complications [42–45].

Within the presented work we are trying to develop a general approach that combines DM and PM techniques to improve simulation procedure by automatic data processing, identification of CPs for diverse patient flow, and integration of predictive models.

## 3 General requirements

Simulation of patient flow is tightly connected with consideration of several aspects to providing valid and realistic simulation. In this section, we collected key identified requirements for such simulation.

R1) Usually, patient flow simulation is focused on the specific category of patients defined by nosology(-ies), particular hospital structure(s), patient's characteristics, etc. Still, the critical problem is the *diversity of the patient flow* in the real world. Within the scope of personalized medicine [1] (in contrast to evidence-based medicine) individual characteristics of patients should be considered. Moreover, clinical processes are characterized by high level of uncertainty [11] which leads to growing variation in clinical episodes. As a result, often, even within a small group of single nosology multiple classes, variations in CPs, and rare but important cases can be identified [10]. To manage this issue PM methodology used to be extended with additional algorithms and solutions. A few examples include algorithms for temporal patterns recognition in CPs [10,27], workflow concept for activity-based PM [23], detailed analysis of conditional branching [26], ontology-based description [46,47], or original algorithms (like dynamic-programming-based algorithm in [48,49]).

R2) Activities simulated within patient flow should follow specific scenarios to *arrange the events* within the natural development of the disease. Typically, the events are arranged within CPs, but to simulate the diversity of patient flow, conditional branching, loops, and interrelated events should be considered explicitly with an extension of appropriate PM procedures to build detailed, realistic and accurate CPs. Such extensions often include DM techniques like decision trees [50–52], the classification for conditional branching [26], prediction of CP development [27], semantic rules [47], fuzzy rules [53] to build detailed, realistic and accurate CPs.

R3) Decisions and activities of *various actors and decision-makers* should be considered concerning conditions and events in clinical pathways. The critical attention of the simulation is around personal activities and experience of either doctors or patient. The interaction between patient and physician is today mainly considered in a context of participatory decision-making [54,55]. Still, to make a comprehensive simulation more complex decision process should be considered. E.g., a set of the interacting actor (usually containing physicians and patients) could be extended with nurses with different specialization [56–58], allied health professionals (including lab technicians, receptionists, etc.) [58,59], visitors [58,60]. Also, complex scenarios of collaboration and interaction during decision making [61,62] could be elaborated.

R4) *Multiple classes of limited resources* are involved in patient's care process including human resources, hospital facilities, wards (with defined capacity), drugs, etc. To make a simulation applicable in various ways a range of issues should be considered: queueing (waiting results, priorities, etc.), scheduling (available periods, access delays, preliminary appointment, staff assignment, etc.), resources constraints (capacity, parallel access, throughput, etc.), access policies (drugs available with prescript, access permissions, required consults, etc.), resources assessment measures (utilization, cost, efficiency, etc.), system-level and stochastic characteristics (complexity, uncertainty, variability, etc.), possible management procedures (adding new resources, rescheduling, queue management, crisis management, etc.). These issues usually get more in-depth consideration within dedicated solutions from operational research and management science areas [63,64] or discrete-event simulation [65–67].

R5) *Multiple levels* [4], *sub-systems* [3], and *sub-processes* [68] should be considered to develop comprehensive simulation solutions that accurately reflect complex healthcare process in hospital [69]. The required level of comprehensiveness could be enabled by generalization and extensibility of the developed approach for modeling simulation and analysis using various approaches: ontology-based systematization of structures and processes [70–72], multiscale modeling and simulation [73,74], or even visualization tools developed for analysis of care process complexity [75,76].

R6) *Multiple data and knowledge sources* can be integrated and analyzed to provide a proper simulation. The area get rapid growth with introduction of big data approaches and technologies [77–79] enabling multiple data sources to be considered for integration: EHR, recommendations of various level (guidelines, standards, laws), population data (census, official reports), omics data, wearable devices, social media, human-generated data (surveys, self-reported data), financial data, pharmaceutical data, and many others. The integration solutions often consider the issues of semantic integration [80,81], managing patient-centered data collections [52,82], building advanced tools for analysis of available data [83,84], as well as general model-based [85,86] and workflow-based [87] integration. These approaches are used to identify models and support simulation (see, e.g., ISPOR Task Force Reports [88]) within different approaches (including model identification, calibration, verification, etc.).

R7) Considering multiple and diverse data sources to be integrated and processed with PM and DM techniques, *data quality issues* become an important problem to be considered during modeling and simulation in medicine and healthcare. Mostly, works in this area are focused on systematizing of qualitative and quantitative metrics, as well as detecting and assessment procedures. Considering EHR as a key data source leads to extended attention to the development of approaches and metrics for assessment and analysis of its quality [89,90]. Case-based studies exploit structural metrics like completeness, timeliness, integrity, validity and other [91]. Model-centered works [85,92] consider data quality in connection with uncertainty and variability analysis, model transparency and validation (this often include discussion on probability or deterministic quality metrics, expert-based quality assessment, incorporation and consistency of data, sensitivity analysis, etc.). On the other hand, PM techniques extend data quality requirements using concepts of cases, activities, events, and attributes assessing correct reflection of reality in the event log (see Section 5.4 in [94]).

## 4 Conceptual framework

Considering the mentioned aspects and issues, a conceptual framework was developed to bring broader but systematic view to the process of building solutions for simulation of patient flow. The developed conceptual framework is presented in Fig. 1 (the figure represents relationships between key concepts which can be directly obtained from sources (blocks with solid borders) or inferred using additional knowledge sources or intelligent technologies (blocks with dashed borders)). It covers the building of the simulation solutions focusing on patient flow within a hospital or set of departments in it. With this consideration, the framework is limited to the corresponding class of solutions. Nevertheless, it was done without loss of generality, and the framework can be easily extended to many other classes of problems and scales. Selected key data sources include EHR, pharmacy and accounting information, schedules of hospital operations which present the major part of information about treatment process and activity within the hospital. All these data sources describe various aspects of the care process. E.g., EHR as a core data source may describe diagnosis (including complications appeared), patient's logistics (departments, responsible physicians, etc.), clinical tests, surgeries and various operations with a patient, and others. Additional data sources describe drugs distribution in hospital (pharmacy), scheduling of human resources (physicians, nurses, etc.) and medical facilities (surgery rooms, imaging facilities, etc.), financial information (including different insurance workflows, patient's payment and, on the other hand, expenses – salary, medication, materials, etc.). All these data may be stored in a single medical information system or several of them. Different data sources may be integrated explicitly (by linking corresponding data), implicitly (by providing enough information for linking), or even have no direct integration (often there are ways to reconstruct links between data using various algorithms).

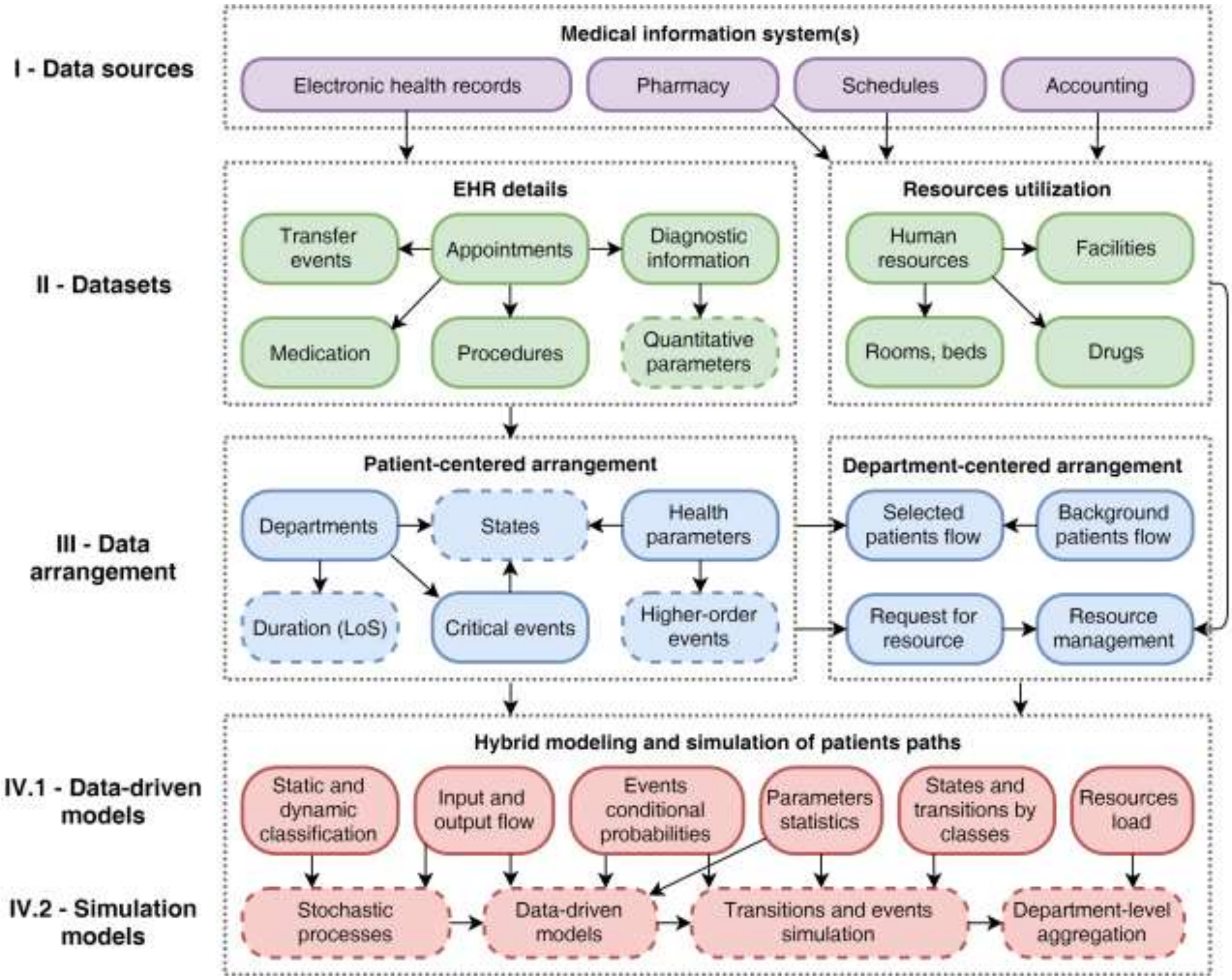

Figure 1 – Conceptual framework for hybrid simulation

The first step considered within the framework is cleaning and *systematization of available data* within a scope of patient treatment (EHR details) and resources utilization (moving from level I to level II of the proposed conceptual framework). These concepts enable checking and correcting of data consistency, detection of missing and contradictory data, obtaining events and attributes inferred with extended knowledge sources and additional intelligence technologies (e.g., retrieving parameters through text and data mining, rule-based inference, etc.). For example, within a scope of patient treatment, linear structure of EHR is extended with links between related events (e.g., appointment of clinical test, taking samples, in-vitro diagnostics), missing events may be reconstructed (e.g. missing outgoing transfer from one department may be identified by presented incoming transfer to another department, missing attributes may be reconstructed from duplicated record (e.g., textual description surgeries, concluding remarks in EHR, etc.)). Additional quantitative parameters may include various metrics on patient state, care quality, risks, etc. Resource utilization scope may include events and attributes reconstruction for particular resources linking these events with corresponding activities of staff (physicians, nurses) involved in the care process.

Next step transforms these datasets into *model-oriented* concepts arranged around patient care and activity of departments (from level II to level III). Patient-centered concepts include states of a patient, transfer between these states with relationship to specific departments of a hospital, high-order events (changing of attributes, knowledge-based detected events, typical sequences of events, and others), critical events with higher significance. This step includes the building of event log, arranged within clinical cases with attributes of events, timestamps, etc. Also, this can be considered as a reconstruction of "personalized" CPs showing everything happened to the patient during the care process. Analysis of this structure can provide higher-level analysis of consistency, completeness, timeliness, etc. For instance, contradictions in events and attributes, disagreements with standards and recommendations can be detected. Important task implemented in this step is an inference of analytical parameters: patient states for each timestamp, LoS in each state or department, high-order events (e.g., changing (growing or lowering) of certain parameters). Taking these parameters into account may significantly increase consistency and completeness of case description. Department-centered concepts include a flow of patients which can be divided into the flow of patients from selected groups and flow of “background” (other) patients. Also, these concepts include resource management where resources of most types usually can be considered as “coming” from external sources and “spent” at the request of the system. E.g., resource utilization, failure rates, overload, etc. can be estimated at this level. Department beds capacity, surgery rooms, working staff can be considered as resources of a different kind. For example, bed occupancy is crucial for effective management of many departments.

Finally, the concepts directly related to simulation of patients can be introduced (from level III to level IV). This set of concepts include a) *data-driven models* which enable to analyze and discover the structure in model-oriented concepts; b) *models for simulation* of patient flow using the discovered structures and relationships. The combination of those two conceptual levels enables detection and incorporation into simulation the diversity of patient flow to capture rare events and emerging phenomena through the simulation. For example, a data-driven model may include classification of patients, DM predictive models for LoS, complications, or outcomes, forecasting of disease development, etc. Whereas, models for simulation may include SD models for disease development, ABM models for personal activities, DES models for CPs simulation and many others.

The developed conceptual framework can be easily extended to consider different tasks and solution. For example, a list of initial data sources can be extended with results of social media crawling, crowdsourcing, participatory decision making, personal wearable devices. Data sets can include various risk assessments, opinions, differential diagnosis, etc. within the additional scope of decision making or insurance, funds, various reports, etc. within the financial/official scope of view. Data arrangement layer may be extended with higher scales of aggregation (hospital, city, population) with integration specific structures (authorities, infrastructure, society, etc.). Finally, each sub-layer in layer IV may be extended with additional types of model, or models of additional aspects (pathophysiological processes, financial processes, human resources, and other). Still, the key issue of the proposed conceptual framework is the demonstration of the basic structure that enables implementation of the simulation solution to answer the proposed requirements through the multi-layered conceptual integration.

## 5 Simulation solution

To implement the simulation solution using the proposed conceptual framework multiple approaches, platforms, and tools can be used. A solution can be developed as a data-driven application for modeling and simulation (e.g., in a distributed data processing environment) [95]. The mentioned ideas can be implemented within a modular medical information system as an extension of the system [96]. Finally, the framework can be considered as a methodology for experimental study [97]. Within this section, a generalized toolbox (see Fig. 2) for building simulation solutions is proposed as a combination of key entities (artifacts in the form of knowledge or data), and procedures to be implemented as well as their interconnections. EHR is considered here as the main data source. Still, a similar approach can be applied with multiple data sources proposed in the conceptual framework.

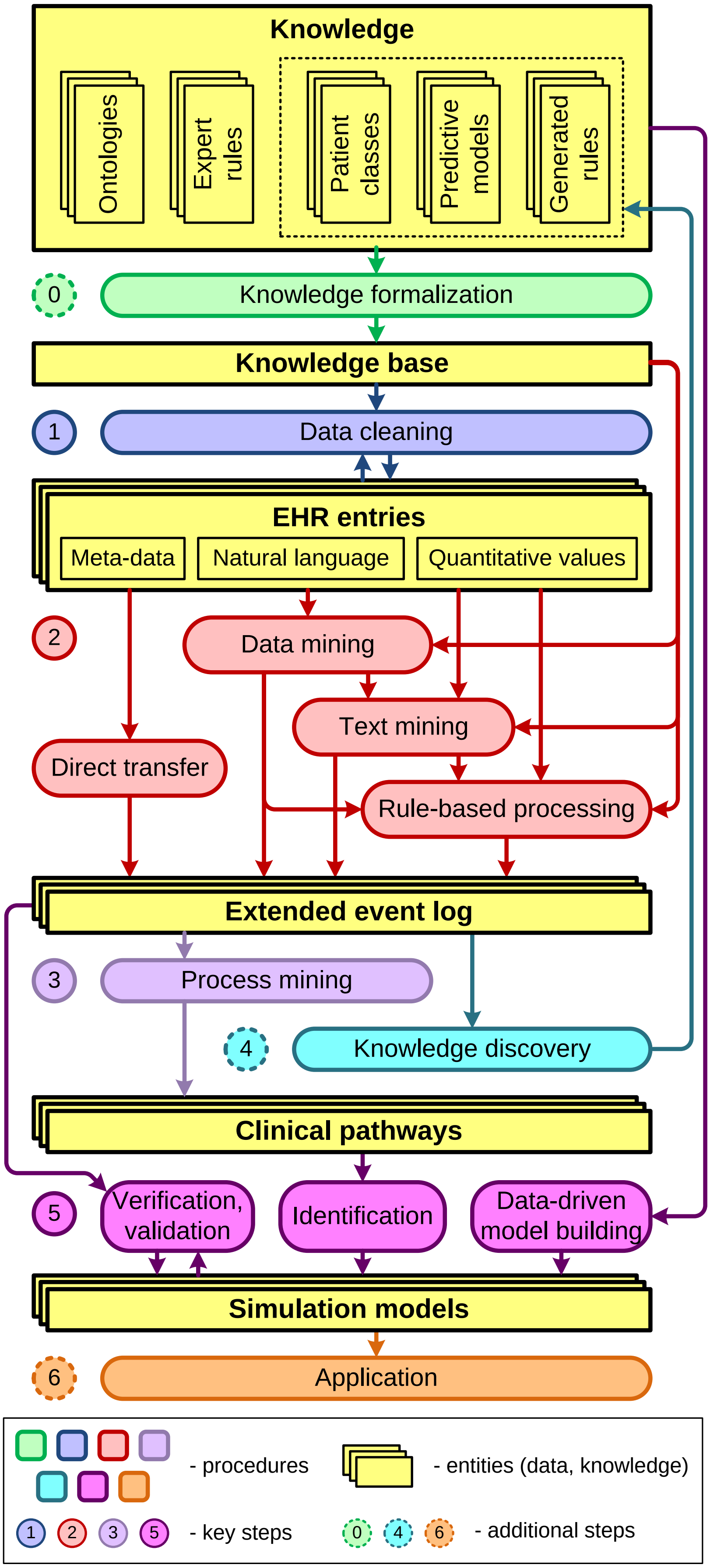

Figure 2 – Toolbox for building hybrid simulation models with EHRs

Procedures within the proposed toolbox are grouped within several steps, forming a workflow of simulation solution building and operating. Key steps are considered as crucial for the proposed approach and define a “backbone” of the most significant parts of the toolbox. At the same time, there are several additional steps, which may also be important, but play supporting role or may introduce some optional extension. Core role within the toolbox is played by knowledge formalized in a knowledge base using various approaches: rules, ontologies, and others (see (0) in Fig. 2 for explicit and dynamic formalization). Nevertheless, the formalization step could be considered as optional within a general scheme under several conditions: the knowledge base could be prepared before implementation of simulation model (external knowledge base), additional formalization and unification may be redundant if single and well formalized knowledge source is used, implicit formalization could combine this step with further steps. The first step of the implementation includes data cleaning (1), which is aimed at structuring, filtering, reconstruction of key datasets from available data sources (partly implementing the transfer from level I to level II in the conceptual framework). The second step (2) includes implementation of data and text mining algorithms as well as rule-based processing to construct unified extended event log which includes events reflecting key processes and attributes arranged around patients and hospital departments. On the next step (3), the extended event log is used to identify and analyze key processes and corresponding CPs. The extended event log may be considered as a combination of levels II and III in the conceptual framework. Here the approach enables identification of events, structure, and attributes even within partly known or (sometimes) unknown processes. The process mining technique [94] provides complete and rapid methods for analysis processes such as process discovery, conformance checking, and identification of bottlenecks. In a context of CPs identification this may be interpreted as identification of correct CPs for selected clinical cases, analysis of descriptive ability of CPs (coverage of events in event log, available cases, etc.), and searching for crucial elements in CPs (critical nodes in CPs for decision making, for resource load analysis, etc.), correspondingly. Indisputably, using process mining software, these methods allow understanding the diversity and complexity of the analyzed process as well as the general relationship of process components. Additionally, analysis of identified CPs enables knowledge discovery (4) as optional steps which introduce possible feedback for enhanced modeling solutions. Such solutions include a) extension of the knowledge base with classification and predictive models, rules, etc. and b) building of models which can be used during the simulation to represents relationships between pathways, events, and attributes. The latter includes the development of data-driven models generated to simulate hidden and unknown processes as well as more complex process structures (conditional branches, loops, etc.). Still, this step is considered as additional (optional), as proper solutions within the proposed general scheme may be developed without it.

Finally, simulation models are implemented (5) as a combination of identified models into a hybrid solution. For example, the solution may be built with the following models (the list can be extended, see level IV in the conceptual framework):

a) Patient flow simulation to reconstruct diverse patients at the entrance of the hybrid model. It can be considered as a stochastic model (e.g., with Monte Carlo methods) generating incoming patients with specific characteristics, inferred from empirical data and expert rules, in defined proportion to reconstruct realistic in-flow of patients.
b) CPs simulation (e.g., using DES) to evaluate the development of the clinical case. This model may include stochastic processes to manage uncertainty, rule-defined steps, and data-driven models for the extended description of the case, or prediction of its development. Still, the core source of information for model identification are CPs, which describe discovered the structure of process being simulated. Patient flow simulation results are used as input data for this model.
c) Random events generators to reconstruct smaller-scale events and specific behavior of patients (in stochastic terms). Implementation of such model (may also be done with Monte Carlo methods) may be integrated into DES models, or data-driven models (for prediction and assessment of clinical episodes).
d) Surrogate models may be used for data-driven reconstruction of unknown relationships or sub-processes, where regular DM and PM is insufficient due to the lack of data or missing structure. These models (usually implemented with machine learning techniques) may be used as a part of other models (a-c) to develop more realistic and comprehensive models.
e) Control models to capture interrelationships between models and data, as well as model management. May be implemented in various forms (including variants like in (c-d) models). However, more essential issues answered by these models include dynamic control of models with data assimilation (improving or switching model using updated EHR, new events registered, updated patient state reports, etc.), altering models or ensemble modeling and simulation, and other high-level modeling and simulation procedures.

The patterns and knowledge discovered previously by data and process mining are used to interconnect and control simulation models to enable more realistic simulation, capturing rare events and variation in patient flow, simulation of flow dynamics identified automatically, etc. Additionally, high-level control of the simulation enables an intelligent combination of clinical pathways with a variation of the events and attributes. Simulation models can be verified using available event log and used in applications within different areas (decision support, training, research, etc.) with a realistic and comprehensive simulation of patient flow. Finally, concluding step (6) may transfer the obtained results to the context of any application (decision support system, training simulation environment, research dashboards).This step is also considered as optional as it may be skipped in many research scenarios where the aim of modeling and simulation is investigation and analysis, while the application solutions (often represented by enterprise-level software intended to be used in practice of healthcare organizations) are not considered as a goal of the project.

## 6 Clinical pathways and flow of patients with acute coronary syndrome

Acute coronary syndrome (ACS) is one of the major causes of death in the world. It is related to more than 2.5 million hospitalizations worldwide every year [98]. Usually, ACS is caused either by myocardial infarctions (MI) or by unstable angina (UA). One of the critical aspects of ACS treatment is early therapy, which usually includes urgent (within 60-120 minutes) or delayed (within 24-72 hours) coronary angiography with possible percutaneous coronary intervention (PCI, i.e., angioplasty, stent placement). One of the key problem regarding ACS treatment is the organization of the required resources within a hospital. The resources include surgery facilities, beds in intensive and regular care wards, human resources, medications, materials, etc. Having highly non-stationary flow of patients with multiscale periodic patterns (day, week, year) a simulation-based solution can serve for various purposes: a) estimation of departments load with different scenarios and dynamics of patient flow; b) analysis of risks of lack or queueing for different types of the resources; c) integration into training solutions or DSS for treatment process optimization.

Within our research, we used a set of 3434 EHRs collected for patients with ACS served in Almazov Centre during 2010-2015. Short description of the dataset is provided in Table 1. For analysis of the patient flow, we considered three types of medical departments involved in ACS patients care. Namely, surgery, intensive care and regular care wards with cardiological specialization which form a group of key departments (GKD) considered for simulation furtherly. Almazov Centre has two principal departments operating in each of these roles and providing care for the patients with ACS 24/7. To avoid the ambiguity of internal identification of the departments here and further departments in GKD for the analyzed dataset are named surgery department (SD) #1 and #2, cardiological department (CD) #1 and #2, intensive care department (IC) #1 and #2. In some cases, admission department (AD) is considered as a part of GKD.

Table 1 – Summary of dataset with ACS patients (3434 patients during 2010-2015)

| Parameter | Value |
|---|---|
| Number of patients per department | |
| - Surgery department #1 | 1716 |
| - Surgery department #2 | 711 |
| - Cardiological department #1 | 2747 |
| - Cardiological department #2 | 144 |
| - Intensive care department #1 | 2012 |
| - Intensive care department #2 | 1762 |
| Origin | |
| - Ambulance | 2531 |
| - From other hospitals | 695 |
| - Other | 214 |
| Patient in-flow (Mean/StDev) | |
| - Patients per month | 47.69 / 18.79 |
| - Patients per day | 1.57 / 1.48 |
| - Patients per hour | 0.07 / 0.26 |
| Sex, male / female | 2476 / 955 |
| Age (Mean/StDev), years | 60.5 / 19.52 |
| Surgery duration (Mean/StDev), min | 43.92 / 28.71 |
| Number of surgeries | |
| - one surgery | 1746 |
| - two and more surgeries | 294 |
| Common complications, % | |
| - Ventricular fibrillation | 5,71 |
| - Cardiogenic shock | 4,4 |
| - Acute pulmonary edema | 7,49 |
| - Pneumonia, pulmonary infiltrate | 4,86 |
| Length of stay (Mean/StDev), hours | |
| - Intensive care #1 | 19.9 / 284.72 |
| - Intensive care #2 | 43.49 / 552.44 |
| - Cardiological department #1 | 107.26 / 498.9 |
| - Cardiological department #2 | 72.74 / 227.82 |
| Outcomes | |
| - Died | 97 |
| - Further treatment | 1799 |
| - Home | 1379 |
| - Other | 159 |

Analysis of the ACS patient flow in GKD reveals diverse and complex patterns. Care process involves multiple departments including multiple visits to departments (possibly different) of the same role. Patients are delivered by ambulance, or transferred from another hospital/department. Each patient has individual properties, complications, treatment plan (surgery, therapy, medication, etc.). There is significant variation in patient in-flow rate (also affected by daily, weekly, yearly patterns) and very high variation in LoS. As a result, complex and diverse patterns of care trajectories can be identified. For example, Table 2 and Fig. 3 show a summary on EHR where the patient went through coronary catheterization and PCI, stayed in different departments (two ICs and two CDs were involved in care process).

Table 2 – Sample EHR summary for single patient (2014)

| Date | Time | Dep. | Event |
|---|---|---|---|
| 1-Jan | 18:43 | AD | Hospital entrance |
| 1-Jan | 19:22 | IC #2 | Department entrance |
| *1 medical test was done* | | | |
| 1-Jan | 19:22 | IC #2 | Check-up by an intensivist |
| *16 medical tests were done* | | | |
| 1-Jan | 23:18 | SD #1 | Coronary catheterization |
| *7 medical tests were done* | | | |
| 2-Jan | 11:34 | IC #2 | Check-up by an intensivist |
| *1 medical test was done* | | | |
| 2-Jan | 11:44 | IC #2 | Check-up by an intensivist |
| *4 medical tests were done* | | | |
| 3-Jan | 10:35 | IC #2 | Check-up by an intensivist |
| *4 medical tests were done* | | | |
| 6-Jan | 14:02 | CD #1 | Check-up by a physician |
| *4 medical tests were done* | | | |
| 9-Jan | 17:15 | CD #1 | Check-up by a physician |
| 10-Jan | 11:30 | CD #1 | Check-up by a physician |
| *1 medical test was done* | | | |
| 13-Jan | 11:23 | CD #2 | Check-up by a physician |
| 14-Jan | 9:58 | CD #2 | Check-up by a physician |
| *8 medical tests were done* | | | |
| 15-Jan | 9:20 | CD #2 | Check-up by a physician |
| *3 medical tests were done* | | | |
| 16-Jan | 9:32 | CD #2 | Check-up by a physician |
| 16-Jan | 13:03 | SD #1 | PCI |
| 16-Jan | 13:39 | IC #1 | Check-up by an intensivist |
| *4 medical tests were done* | | | |
| 17-Jan | 9:16 | CD #1 | Department entrance |
| 17-Jan | 11:38 | CD #1 | Check-up by a physician |
| *1 medical test was done* | | | |
| 20-Jan | 9:54 | CD #1 | Check-up by a physician |
| 21-Jan | 10:43 | CD #1 | Check-up by a physician |
| 22-Jan | 9:47 | CD #1 | Check-up by a physician |
| 23-Jan | 9:21 | CD #1 | Check-up by a physician |
| 24-Jan | 11:57 | CD #1 | Discharge from hospital |

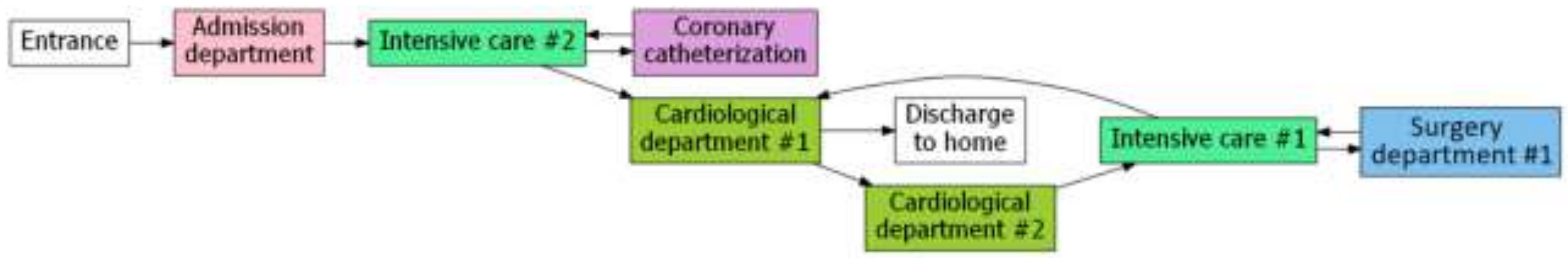

Figure 3 – Sample EHR trajectory within GKD

Moreover, to analyze and simulate a flow of patients with ACS within a scope of both patients and departments a full flow of patients through the GKD should be considered with division into two parts. The parts include the flow of patients with ACS and the flow of other patients (furtherly, "background flow"), which has even higher diversity and more complex patterns. In our case background flow influences the target flow by occupying the same surgery facilities, beds in the departments, providing an additional load for departments' staff. E.g., Almazov Centre performs planning stenting procedure for patients from all around Russia. Fig. 4 shows the major transfers between the departments (transfer probability higher than 5% are shown, green lines depict inter-departments transfers with the probability higher than 20%, orange lines shows transfers from and to external sources) for background flow in GKD.

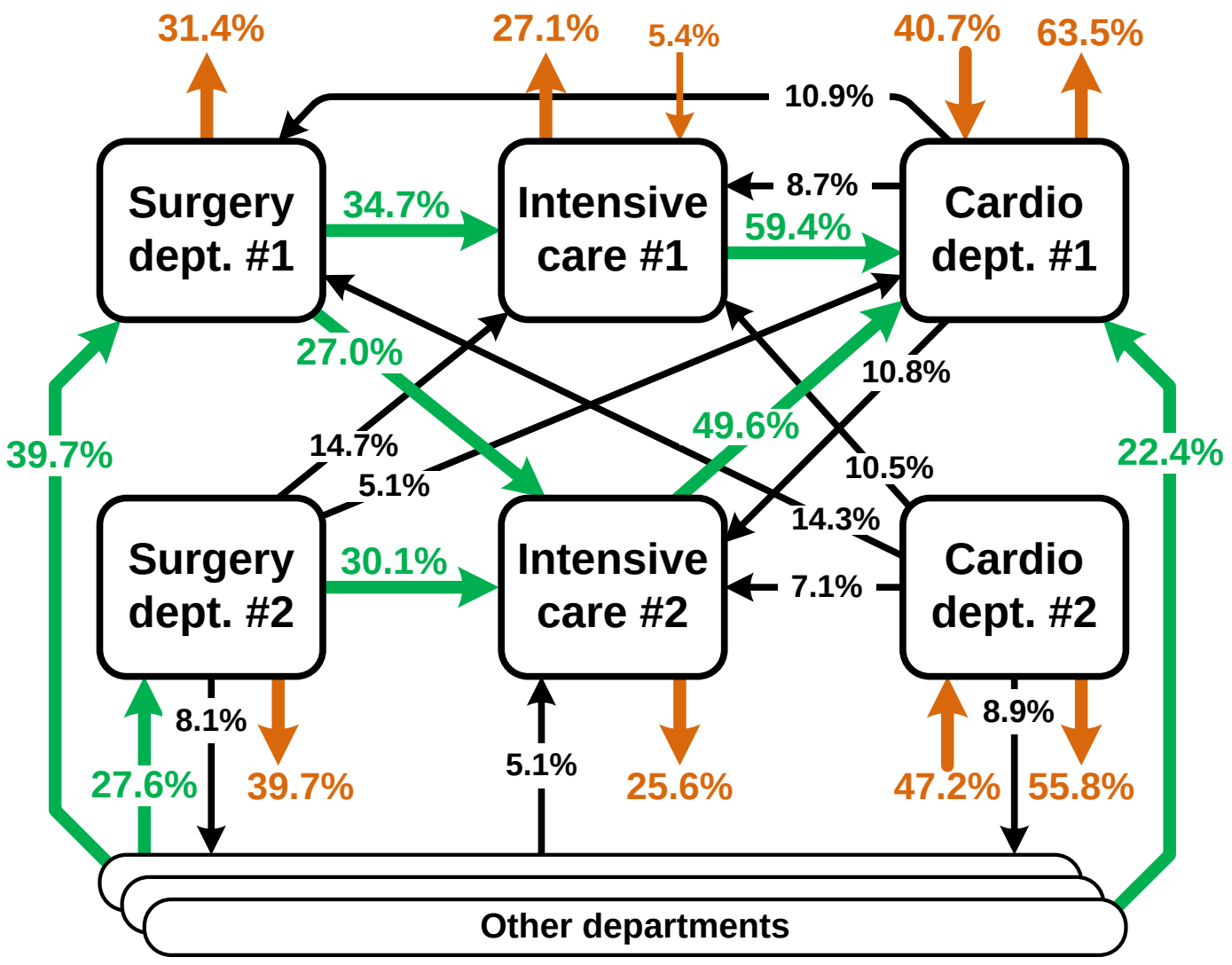

Figure 4 – Patient flow within a group of key departments (background flow)

The approach proposed earlier in Sections 4-5 was applied to identify and simulate the patient's behavior and the load of the GKD within ACS treatment to analyze the complex patterns of the patient flow in more details.

### 6.1 Identification of complex processes

Almazov Centre uses medical information system qMS[2] which support exporting EHRs into XML documents. Each EHR consists of multiple events such as an entrance, tests, checkups by physicians, surgeries, discharge, etc. An event is described by date, time, a place, a title, and staff who executed it. Event data can include extended information on anamnesis and diagnosis, test results, description of procedures, etc. Still, EHRs often suffer lack consistency, completeness and correctness due to the weak structure (including plain text), missing data (hand-written paper documents are still in practice), technical problems (synchronization, consistency checking, etc.) and many other problems [99]. For data cleaning purposes (Step #1 of the toolbox presented in Section 5) we perform basic heuristics including sorting of EHR events (applied to 89.3% of records) and reconstruction of missing events (applied to 11% of records) to clean the data and improve its consistency.

[2] https://sparm.com/products/Qms (in Russian)

The next step (Step #2 of the toolbox) include structuring and encoding (so the sequence of states $S_i$ is represented by a string of letters of length in a range 1..59 characters) patient's' CPs. The encoding is done concerning the departments and sub-departments (facilities with particular specialization within a department) related to the current state of the patient as well as cyclic patterns repeated in multiple pathways. For instance, the sequence AFIFE means the patient firstly admitted to hospital (letter A is for admission department), then he/she moves to intensive care unit (letter F), then to operating room to get PCI (letter I), again to IC (letter F) and finally to cardiological department (letter E). The encoded sequences can be clustered to identify sustainable groups of patients. Using K-means method with the Levenshtein distance, we got 13 clusters which were identified as the best number of clusters according to cluster coefficients of variation [100] which is the ratio between the intracluster coefficient of variation and the intercluster coefficient of variation. The less this ratio is, the better number of clusters is.

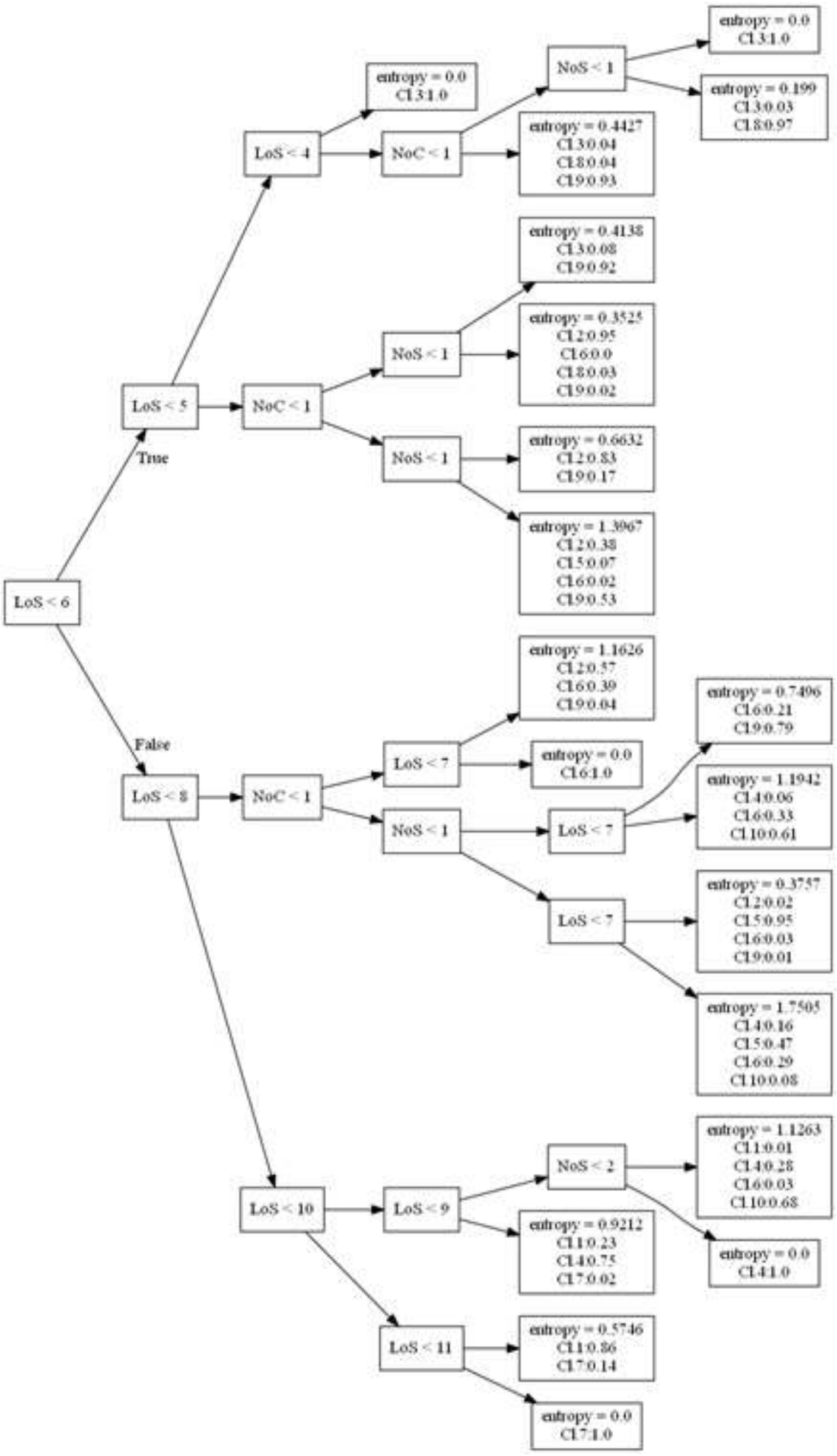

Figure 5 – Reduced decision tree to identify clusters of patients

Among the clusters, three of them include less than five sequences, and these clusters were discarded. For a deeper analysis of clustering structure, a decision tree was built based on the CART algorithm using scikit-learn[3] (Fig. 5). The decision tree shows the distinction between clusters and helps for further simulation of CPs. Internal nodes test one of three possible attributes of the patient's moving sequence (LoS – a length of the sequence, NoC – a number of coronary catheterization, NoS – a number of surgeries). The branches represent the outcome of the tests (an upper branch means an attribute is tested positively and a lower branch means the opposite). Leaf nodes represent the probabilities for possible clusters according to this classification path in the tree. Here clustering and building the decision tree may refer to the loop within the toolbox proposed earlier in Section 5 through the Steps #4 and #0 of the toolbox (see patient classes and predictive models in Fig. 2).

[3] http://scikit-learn.org

After dividing patients into groups, it needs to match a template (i.e., typical CPs after Step #3 of the toolbox) for alignment of events sequences within a cluster. To match better template for a cluster its structure should be analyzed: most common number of surgeries for this cluster; the last department before the discharge (cardiological department or other departments), etc. To identify more peculiarities for a cluster, it is possible to use multiple alignment algorithms from bioinformatics which usually used to align three or more biological sequences [101]. We used an original alignment algorithm which identifies a location of a state in a patient flow compared with a template. For instance, the template of cluster #2 is AEFNINFEDE and sequences AFIFDE, AFINFE, and AFIFD are in the cluster #2. According to the template, sequences are aligned like A0F2I4F6D8E9, A0F2I4N5F6E7, and A0F2I4F6D8 where numbers mean the location of a department in the template. The last step of identification of typical pathways for detected clusters is to form a flow with alignment sequences.

```
S ← {S_1, S_2, ..., S_n}  set of sequences of states
krange ← range for clusters' numbers to select the best one

kbest ← ClusterAnalysis(Kmeans, S, krange)   ▷ define the best number of clusters
clusters ← Kmeans(S, kbest)
for i from 1 to kbest do
    cluster ← clusters_i
    expertparameters ← parameters to define a template
    template ← an appropriate template according to expertparameters
    aligned ← Align(cluster, template)
    ShowCP(aligned)
```

Algorithm 1 – K-means clustering and CP's discovering through expert templates

Fig. 6-7 depict typical CPs which are represented by directed graphs with edges weighted according to flow (number of patients moved this edge within the detected path in a graph) for ten obtained clusters. The edges assigned a flow for 70 % of cluster's patients are shown in bold to depict the most common pathways in the cluster. Presented information obtained in an automatic and semi-automatic way can be effectively used furtherly to reconstruct patient flow within a simulation solution (see Section 7). For a deeper analysis of the clusters and obtained CPs, medical interpretation of the clusters' CPs is provided in the next sub-section.

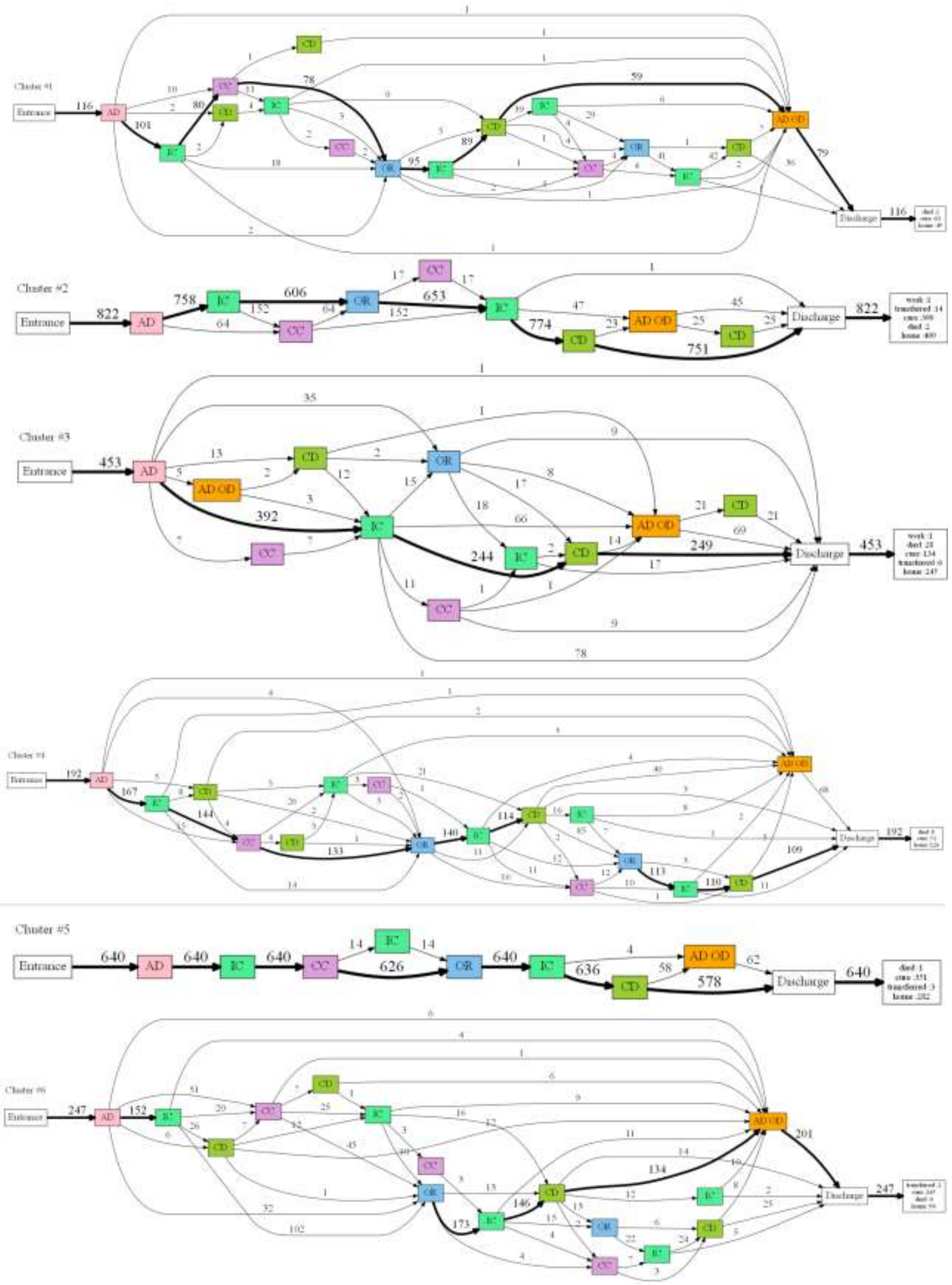

Figure 6 – Patients' typical pathways for clusters ##1-6 (legend is provided in Fig. 7)

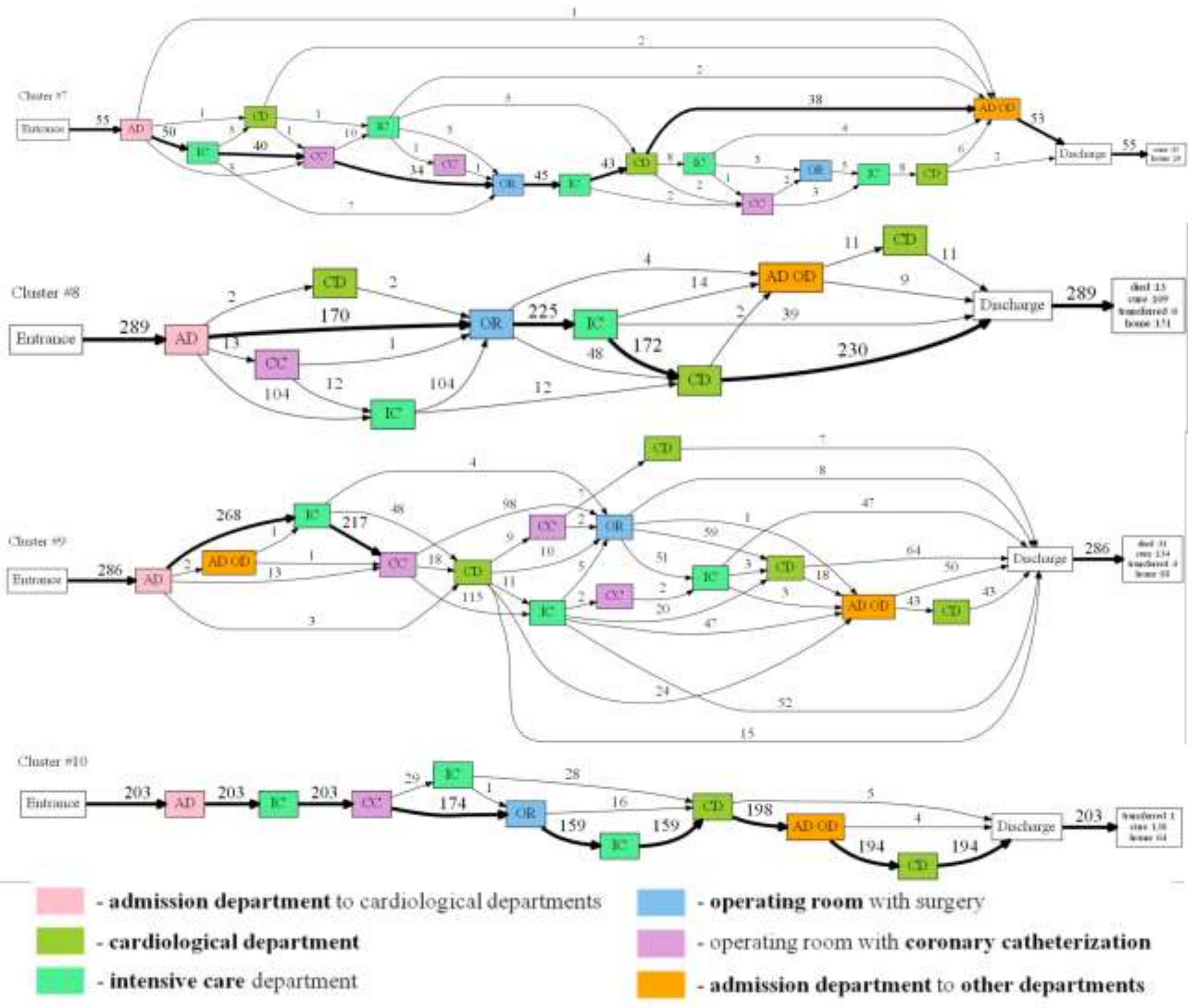

Figure 7 – Patients' typical pathways for clusters ##7-10

### 6.2 Clinical interpretation of clusters

Medical examiners collect the results of randomized controlled trials and other reliable evidence and present them as clinical recommendations to draw up federal and regional regulatory documents which called protocols. The protocols determine the most effective interventions and treatment for fixed groups of patients which are also defined according to these documents. However, protocols are incarnated by different hospitals differently. Each hospital has its unique set of departments, a number of beds and staff structure. Nevertheless, main steps of medical care are identical in medical centers, especially for the most common diseases.

At the same time, randomized controlled trials are carried out with specially selected cohorts of patients. Older adults, patients with atypical forms of diseases or with severe concomitant diseases and comorbidities are usually excluded from randomized trials. However, such groups of patients are most in need of medical care. Moreover, treatment and care strategies are defined by a physician according to patient's peculiarities and personal medical experience of the doctor. The protocol defines the way of treatment in general and does not determine how to care for each patient.

Based on sequences of medical events, clustering of patients is an analytic approach that goes beyond a nosology approach. Patients with various forms of a disease can be cured with the same treatment. On the other hand, peculiar patients with identical forms of the disease have different strategies for treatment. Such an approach of cure is often presented in emergency departments for patients entranced with ACS. For instance, a cohort of patients without ST elevation can be cured with PCI as patients with ST-elevation and have positive treatment outcomes [102].

In Fig. 6 the ***Cluster #5*** contains patients who have had PCI and their treatment strategy agree with clinical recommendations in the best way [103,104]. At the same time, there a group of patients inside the cluster which stay in IC department between coronary catheterization (CC) and PCI in operating room. These patients need additional time before the surgery to get additional medical examinations and doctors' consultations or to discuss the necessity of the intervention with their relatives. Also, a small group of patients from the ***Cluster #5*** move from IC department to other non-cardiological departments because they have coronary or neurological complications or need additional examinations to diagnose acute comorbidities. The ***Cluster #10*** are similar to the ***Cluster #5***. Most patients of the ***Cluster #10*** move from CD to other departments for lengthy rehabilitation.

In Fig. 6-7 many other clusters contain the same step with other departments (AD OD), but each of them has its peculiarities. Namely, patients from the ***Cluster #1*** have repeated PCIs. Patients from the ***Cluster #4*** are cured conservatively without surgeries. ***Cluster #6*** consists of patients who need many complicated medical examinations before PCI. Finally, patients from ***Cluster #7*** stay long in different departments to prepare additionally before the surgery.

The most massive ***Cluster #2*** presents the most optimal strategy of treatment for patients delivered from reference hospitals to perform PCI. Moreover, the ***Cluster #2*** can contain patients with staged PCI. These patients need several hospitalizations to cure their disorder [105]. The ***Cluster #8*** also contain patients delivered from other hospitals like the ***Cluster #2***. However, these patients are in an unstable state and need urgent surgery. Also, the ***Clusters #8*** presents people with myocardial infarction who were delivered with an ambulance in a state of a cardiogenic shock or a clinical death.

The ***Cluster #3*** and the ***Cluster #9*** have a high rate of hospital lethality and show many treatment strategies including conservative ones. The ***Cluster #3*** contains more patients than another cluster and present patients with severe concomitant diseases and comorbidities, atypical diabetes and the primary disease with a severe course. These patients underwent myocardial revascularization which was belated because of contraindications, technical impossibility or ineffective palliative revascularization. The ***Cluster #9*** contain patients with the same problems and have the highest death rate. Also, the treatment strategies are various in this cluster. Most of them include CC, and only a part of them include PCI in OR. The high rate of patients with CC suggests that doctors could access the risk of the intervention.

To sum up, the cluster which has mostly agreed with clinical recommendations sequence of events is not the largest one. Many clusters have the same treatment step with other departments (AD OT) to provide patients with additional rehabilitation which is widely available in Almazov Centre. The clusters with the highest death rate contain many varieties of treatment strategies. The worse outcomes are inherent in patients with belated coronary catheterization and percutaneous coronary intervention.

# 7 Simulation of patient flow

## 7.1 Simulation implementation

A solution was developed (Step #5 of the toolbox presented in Section 5) using discrete event simulation (DES) approach and queuing theory to simulate identified patient flow (release of the developed solution [106] was published at Zenodo [107]). To implement the solution, we used SimPy[4] library for Python which implements basic concepts and enables easy development of DES solutions using basic concepts of processes, shared resources, events tracked within the simulation environment. The basic architecture of implemented DES solution is presented in Fig. 8. The processes within the solution include two generators for background/target patient flows and set of patients (each represented by a single process).

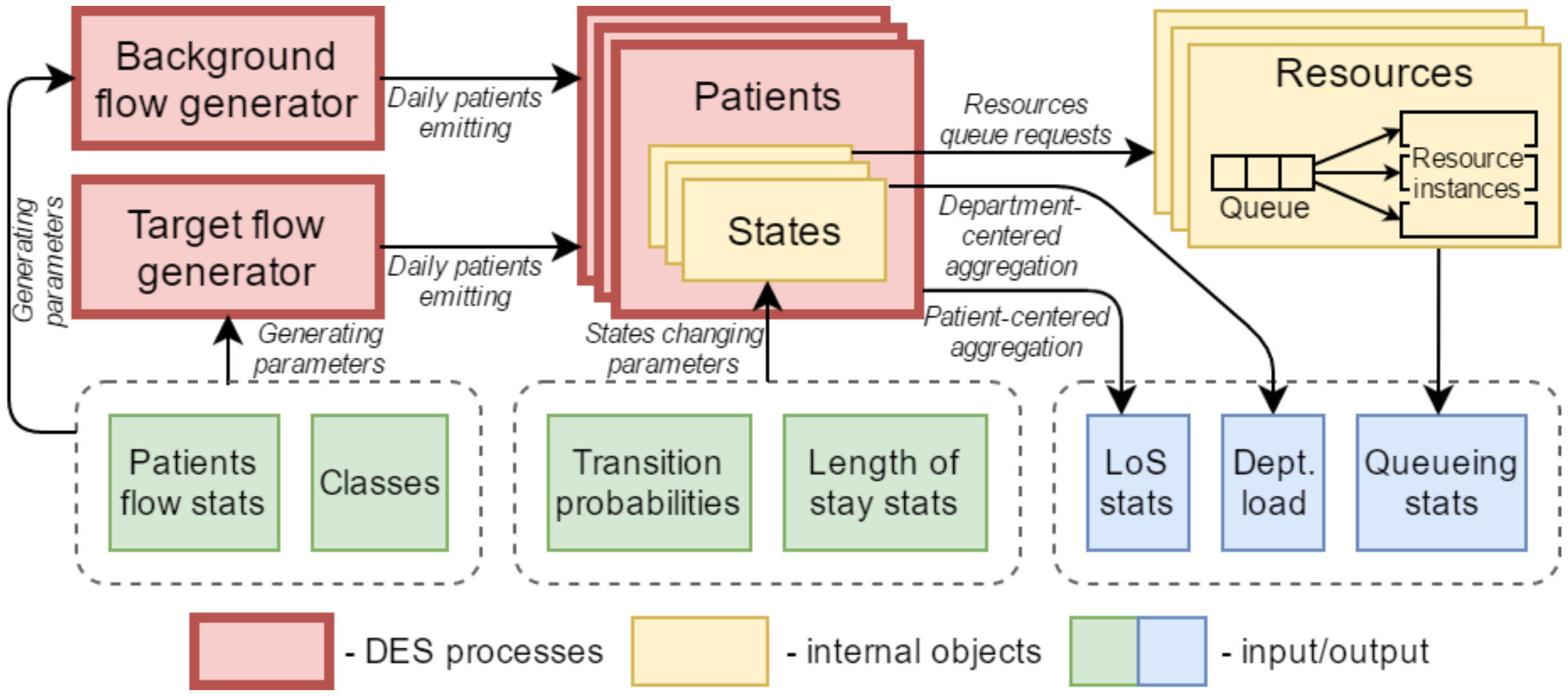

Figure 8 – Simulation implementation

Flow generators produce a sequence of incoming patients in initial states with defined class and arrival time. To generate those sequences empirical distributions obtained from data are used.

[4] http://simpy.readthedocs.org/

Fig. 9 presents PDFs for simulation of daily ACS patient flow (Fig. 9a,b) and random selection of patients class according (Fig. 9c). To generate random sequences SciPy[5] library classes were used. The daily in-flow sequence is generated at the beginning of each simulated day and includes timestamps for patients' arrivals. Two generators (background and target) use corresponding in-flow sequences and iterates between timestamps with event generation after a required time delay. Each event generates and initiates processing of a patient instance. For target flow CP class is selected during generation.

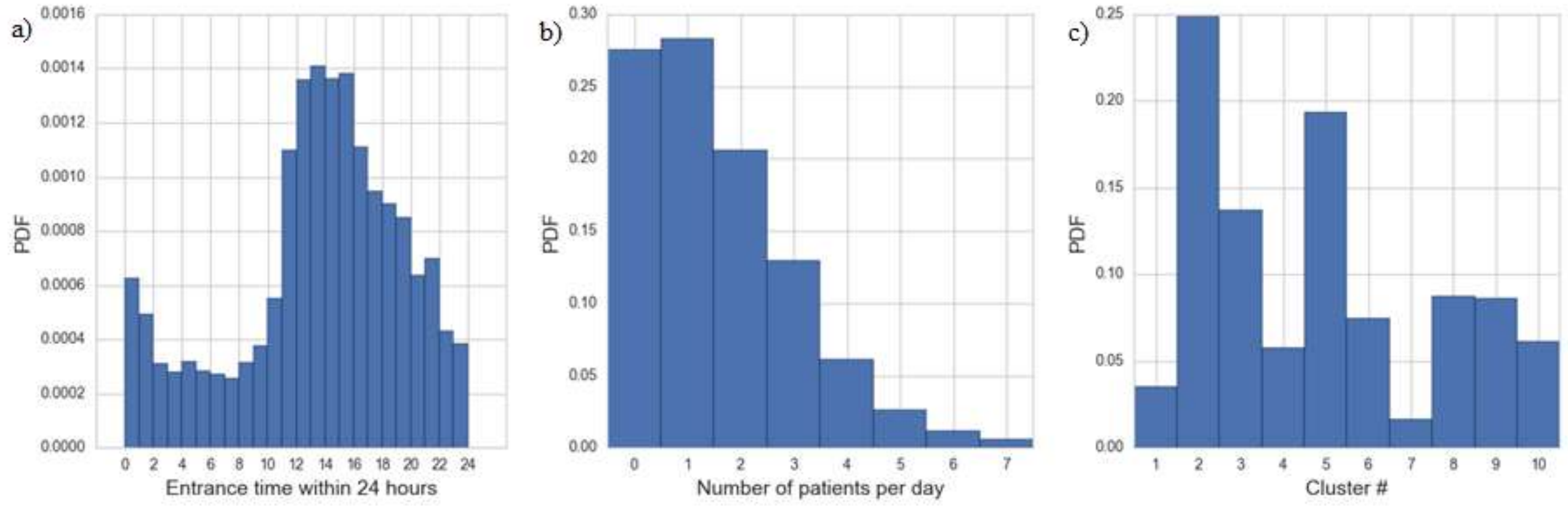

Figure 9 – Probability distributions for ACS patient flow: a) entrance time within 24 hours; b) number of patients per day; c) cluster ID

Each generated patient is presented by a process instance which controls the transfer of the patients between states by transition probabilities and LoS distributions associated with the selected cluster (see Section 5.1). Transfer probabilities are defined by between a matrix with all states mentioned in corresponding CP.

Selected states related to usage of limited resources with possible queueing (e.g., surgery facilities). Simulation of switching to such states is implemented within requesting and waiting for access to one of a predefined number of resources of the particular type. For instance, within the detected pathways CC and OR states are related to queue simulation as it requires access to angiography equipment.

To provide a stochastic simulation of selected scenario multiple runs of the simulation is performed. To make experimental study faster JobLib[6] was used to run independent simulations in parallel mode on a multi-processor environment.

Within the implemented experimental solution, we used two input parameters: number of available resources (here, surgery rooms) and in-flow scaling rate. Simulation results include statistics on generated process instances (e.g., total LoS, LoS for each state) and resource usage (e.g., queueing time, number of cases experienced waiting in the queue, number of patients in each state at each moment). These two group of parameters may refer to patient-centered and department-centered consideration.

[5] http://scipy.org/
[6] http://pythonhosted.org/joblib/

### 7.2 Simulation results and discussion

To test implemented approaches flow of ACS and background patients with the detected classes of CPs were simulated. To analyze the capabilities of the proposed approaches, we implemented a baseline solution which uses no information on CPs and clustering and relies only on empirical transition probabilities between basic states (namely, AD, CD, IC, OR, CC, AD OD). The analysis of the simulation results shows that the proposed approach (implemented as a combination of CPs classification and DES simulation) enable improvement of simulation to generate more realistic patient flow. Fig. 10a shows QQ-plot for LoS obtained with the proposed approach for selected top 3 clusters (namely ##2, 5, 3 which covers 58% of all the cases) in comparison to baseline solution ('No classes'). Kolmogorov-Smirnov statistics decreased by 51% (from 0.255 to 0.124). More accurate control of patient flow provides the better fitting for patients with a longer period of stay (more than 20 days) which is important to simulate rare but complicated cases.

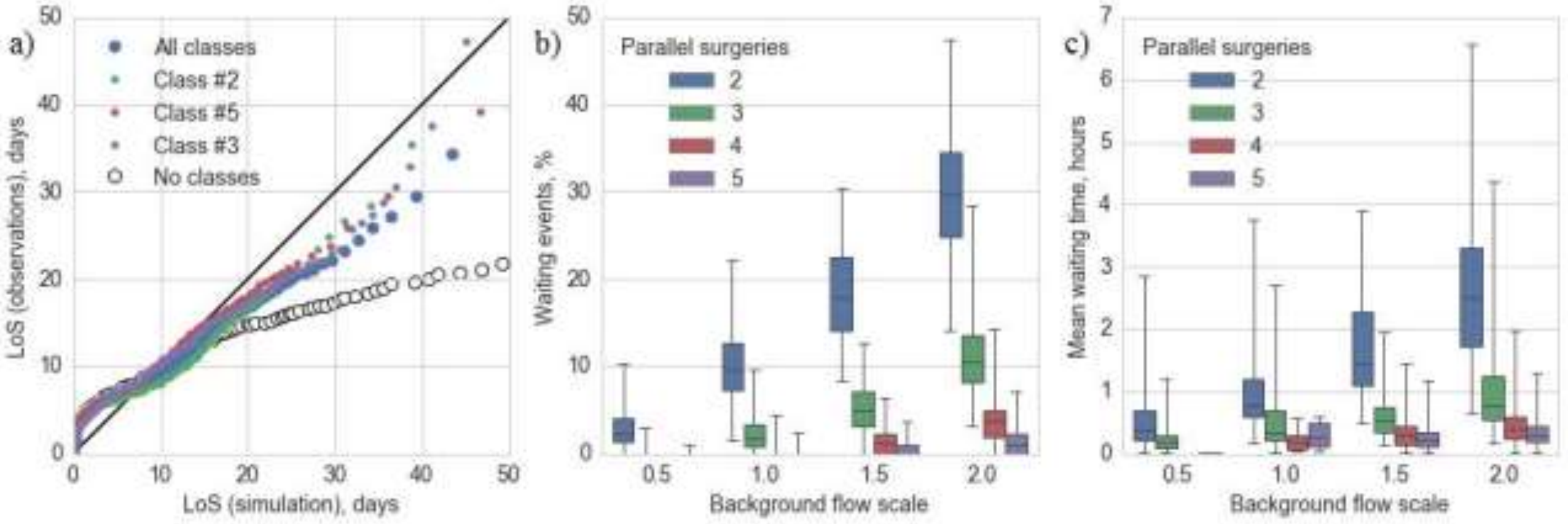

Figure 10 – Simulation evaluation a) QQ-plot for LoS in simulation and observation; b) percentage of cases which experienced queueing; c) average waiting time for simulation period

For the demonstration of possible further analysis and application (Step #6 of the toolbox presented in Section 5), we perform a simulation of 60 working days. The simulation was performed with a different number of angiography equipment available, and scaling of patient flows to investigate the influence of removing (e.g., due to failure) or adding angiography facilities in the hospital. Simulation we performed for 2 to 5 parallel surgery facilities available and for scaling of background patient flow with rates 0.5, 1.5, 2.0 (1.0 is default scale identified from empirical data). Each simulation was performed 100 times to analyze the stochastic behavior of the model. Fig. 10b and 10c show percentage of simulated patients which experienced waiting in a queue and average queueing time. Each boxplot shows distribution (minimum, Q1, Q2, Q3, maximum) of the mentioned parameter in 100 experiments. For example, analysis of such results may show that in our experimental case switching from 3 to 4 parallel surgeries available may significantly decrease the risk of waiting in a queue. For default scale, this changes all quartiles to 0%. Also, average waiting time is declining. Maximum of 100 experiments is less than 1 hour, while the maximum for 3 parallel surgeries almost reached 3 hours.

The results of such simulation can be used in decision support and policy-making to improve patients serving within different conditions with stochastic estimation of waiting time and thus assessing risks of complications and death after PCI as well as departments and facilities load. The conditions may include internal structure of the hospital (available surgery facilities, the schedule for facilities and department, etc.) and external parameters (e.g., the variability of patient flow, diagnosis, etc.). Few examples of the solution application include the following scenarios:

- what-if analysis or optimization for selection of best possible facilities' capacity to lower risk for incoming patients systematically;
- investigation of various resources utilization (including surgery facilities, human resources, and other), analysis of the provided care quality and cost;
- stochastic prediction of hospitals, departments or facilities load for planning and management purposes;
- development of simulation solution for the training of physicians, nurses of the hospital within various scenarios;
- enhancement of internal hospital rules and recommendation for patient care;
- planning "on-demand" extra facilities obtained by collaboration between departments of a hospital or even between hospitals.

Also, it worth to mention that the proposed solution can be easily extended in many directions to enhance its functional characteristics. Several examples include more diseases, complications, and comorbidities; events and parameters within CP; DM and PM techniques; machine learning sub-models.

## 8 Conclusion and future works

The proposed approach is aimed towards the extension of simulation solutions with a combination of detailed data analysis for identification of CPs and further simulation of patient flow. A combination of data, text, and process mining techniques are used to detect and assess diversity in patient flow, structure, and classes of CPs, etc. As a result, the approach enables automatic identification of patients' dynamics on micro-level to perform more realistic simulation and obtaining macro-level characteristics as department load, queueing parameters, patients experience, etc. The demonstrated example shows an implementation of the proposed approach to improve discrete-event simulation of ACS patient flow using automatic identification and classification of CPs.

Further development of the proposed approach includes the following directions: additional systematization an extension of the conceptual framework for consideration of place complexity [108], integration with predictive models [31] to simulate the behavioral and clinical evolution of cases, integration with quality of life metrics [109] to enable detailed analysis of patients experience, and others. On the other hand, future works on the developed implementation of the proposed approach include: generalization of the solution (to build a framework for hybrid simulation of patient flow) with detailed usage of CPs with multiple level of events and attributes detected with process mining approaches, implementation of surrogate models which can be identified and used to make the simulation more detailed and realistic, use modeling and simulation to reconstruct missing and erroneous data, and others. Finally, the developed approach gives an ability to deal with different patient data in an automatic or semi-automatic manner. Therefore, one of the critical direction of further development is the implementation of a data-driven solution to cope with large and diverse datasets using, e.g., by the use of corresponding big data and cloud computing concepts [95].

***Acknowledgments.*** This research is financially supported by The Russian Scientific Foundation, Agreement #14-11-00823 (15.07.2014).